\documentclass[10pt]{extarticle}

\usepackage{nicefrac}

\usepackage[bordercolor=white,backgroundcolor=gray!30,linecolor=black,colorinlistoftodos, textsize=10pt]{todonotes}

\usepackage{amssymb}
\usepackage{relsize}
\usepackage{mathtools}
\usepackage{textcomp}
\usepackage{amsmath}
\usepackage[textstyle,amssymb]{SIunits}
\usepackage{graphicx}
\usepackage{fixltx2e}  %for textsubscript
\usepackage[normalem]{ulem}  % for underline allowing linebreaks
\usepackage{soul}
\usepackage[twocolumn,textwidth=183mm,columnsep=6mm,top=20mm,bottom=25mm]{geometry}
\usepackage{helvet}
\usepackage{titlesec}
\titleformat*{\section}{\bfseries\sffamily}
\titlespacing{\section}{0pt}{*4}{*0}
\titleformat{\subsection}[runin]{\normalfont\bfseries}{\thesubsection.}{3pt}{}
\usepackage{setspace}
\usepackage[breaklinks=true]{hyperref} %[breaklinks=true]
\usepackage[hang,flushmargin]{footmisc} 
\usepackage[font={footnotesize,sf},labelfont={sf,bf},labelsep=endash,justification=raggedright]{caption}

\DeclarePairedDelimiter\abs{\lvert}{\rvert}%
\usepackage{natbib}
\newcommand\blfootnote[1]{%
  \begingroup
  \renewcommand\thefootnote{}\footnote{\noindent#1}%
  \addtocounter{footnote}{-1}%
  \endgroup
}

\usepackage[labelfont=bf,justification=justified]{caption}
\newcommand{\highlight}[1]{%
	\colorbox{black!10}{$\displaystyle#1$}}

\usepackage{footnote}
\usepackage{footmisc}
\renewcommand{\thefootnote}{\fnsymbol{footnote}}
\renewcommand{\thefootnote}{\alph{footnote}}

\begin{document}

%\title{Theory of frequency modulated combs induced by spatial hole burning, dispersion and Kerr}
%\title{Optical frequency comb in a Bloch oscillator}
%\title{Giant Kerr nonlinearity induced by Bloch gain in quantum cascade lasers and its role in frequency comb generation}
%\title{Frequency comb generation by a Bloch gain induced giant Kerr nonlinearity}

\twocolumn[\begin{@twocolumnfalse}
	%******************* title *******************
	{\centering\LARGE\sf \textbf{\noindent Frequency comb generation by Bloch gain induced giant Kerr nonlinearity}}
	\vspace{0.4cm}
	
	%******************* authors *******************
	{\sf\large \textbf {Nikola~Opa\v{c}ak$^{*}$, Sandro~Dal~Cin, Johannes~Hillbrand, Benedikt~Schwarz$^{\dagger}$}}
		\vspace{0.5cm}
		
		{\sf \textbf{Institute of Solid State Electronics, TU Wien, Gu{\ss}hausstra{\ss}e 25-25a, 1040 Vienna, Austria
}}

		\vspace{0.5cm}
\end{@twocolumnfalse}]
\vspace{0.5cm}

%instantaneous frequency chirp is favored due to di Dispersion leads to a chirp and spatial hole burning cause chirped frequency modulated comb generation}

%\author{Nikola~\surname{Opa\v{c}ak}}
%\email{nikola.opacak@tuwien.ac.at}
%\author{Sandro~\surname{Dal Cin}}
%\author{Johannes~\surname{Hillbrand}}
%\author{Benedikt~\surname{Schwarz}}
%\email{benedikt.schwarz@tuwien.ac.at}
%\affiliation{Institute of Solid State Electronics, TU Wien, Gusshausstrasse 25-25a, 1040 Vienna, Austria}

%\affiliation{Institute of Solid State Electronics, TU Wien, 1040 Vienna, Austria}

%word count
%figure 300/(0.5*1260/561.)+40 = 300
%figures about 600 words
%math 16*lines
%16*(3+12+13)=450

%******************* abstract *******************
{\noindent\sf \small \textbf{\boldmath
		\noindent Optical nonlinearities are known to provide a coherent coupling between the amplitude and phase of the light, which can result in the formation of periodic waveforms. Lasers that emit such waveforms are referred to as optical frequency combs. 
        Here we show that Bloch gain -- a nonclassical phenomenon that was first predicted in the 1930s -- plays an essential role in comb formation in quantum cascade lasers (QCLs).
        We develop a self-consistent theoretical model that considers all aspects of comb formation: bandstructure, electron transport, and cavity dynamics. 
        It reveals that Bloch gain gives rise to a giant Kerr nonlinearity and serves as the physical origin of the linewidth enhancement factor in QCLs.
        Using a master equation approach, we explain how frequency modulated combs can be produced in Fabry-P\'{e}rot QCLs over the entire bias range. In ring resonators, Bloch gain triggers phase turbulence and the formation of soliton-like patterns.
	}
}
\noindent
\blfootnote{\noindent$^*${nikola.opacak@tuwien.ac.at}, $^\dagger${benedikt.schwarz@tuwien.ac.at}}

\begin{figure*}[ht]
	\centering
	\includegraphics[width = 1\textwidth]{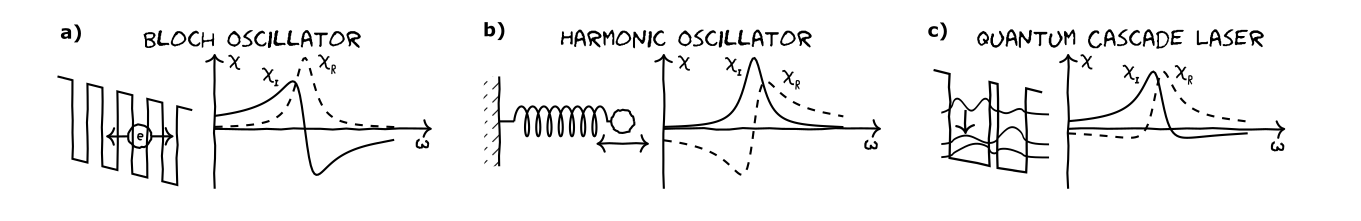}
	\caption{ \textbf{Illustration of different systems and their optical spectral response.}  \textbf{(a)} Nonclassical Bloch oscillator in a semiconductor superlattice and a \textbf{(b)} classical charged harmonic oscillator and their complex susceptibilities $\chi=\chi_R+\chi_I$. Optical gain, which is proportional to $\chi_I$, has a symmetric Lorentzian shape and $\chi_R$ has a dispersive shape in the case of the harmonic oscillator. For a Bloch oscillator, the shape profiles of $\chi_R$ and $\chi_I$ are exchanged compared to the harmonic oscillator, due to a $\pi/2$ phase shift in the spectral response. This results in the dispersive shape of the Bloch gain. \textbf{(c)} Schematic of the QCL bandstructure with the laser levels and the optical transition. The complex susceptibility can be represented as a sum of both Bloch and harmonic contributions.}
	\label{fig_intro}
\end{figure*}

Bloch and Zener predicted charge oscillations in a periodic potential under an applied constant electric field in the 1930s~\cite{bloch1929ueber,zener1934theory}, a phenomenon which is commonly referred to as Bloch oscillations.
%Oscillatory motion of charge carriers in a periodic potential under an applied electric field was first predicted by Bloch~\cite{bloch1929ueber} and Zener~\cite{zener1934theory} in 1929 and 1934 respectively, and are commonly referred to as Bloch oscillations. 
It attracted researchers ever since due to the property of oscillating charges to couple with electromagnetic waves, potentially offering new sources of radiation~\cite{waschke1993coherent}. In condensed-matter theory, the motion of electrons in a periodic crystal lattice is governed by the energy-momentum relation within a Brillouin zone~\cite{ashcroft1976solid}. A constant electric field accelerates the electrons towards the edge of the Brillouin zone, where they experience Bragg reflection, resulting in an oscillatory motion. The width of the Brillouin zone in bulk crystals is large and thus electrons cannot reach the edge before they scatter. However, in semiconductor superlattices (Fig.~\ref{fig_intro}a), made of alternate semiconductor layers~\cite{esaki1970superlattice}, it is significantly narrower and electrons can complete multiple oscillation cycles within their lifetime~\cite{leo1992observation,feldmann1992optical}. Ktitorov et al.~\cite{ktitorov1971bragg} %and Ignatov et al.~\cite{ignatov1976nonlinear} 
predicted tunable optical Bloch gain arising from these oscillations, which was subsequently verified in a GaAs/AlGaAs superlattice~\cite{sekine2005dispersive}. 
The gain is present even without population inversion, a necessary ingredient in the analysis of a classical harmonic oscillator (Fig.~\ref{fig_intro}b). 
%an ingredient that is an imperative to obtain light amplification~\cite{siegman1986lasers} in the analysis of a classical harmonic oscillator (Fig.\ref{fig_intro}b).
Moreover, the Bloch gain possesses an S-shaped profile (Fig.~\ref{fig_intro}a), referred to as the dispersive gain~\cite{sekine2005dispersive}.
%due to its non-classical nature . 
This unique spectral response, sharply contrasted with the well-known symmetric Lorentzian gain of a harmonic oscillator (Fig.~\ref{fig_intro}b), serves as the fingerprint feature of the Bloch gain.
% In sharp contrast to well known symmetric Lorentzian gain shape of the classical harmonic oscillator, Bloch gain is characterized by its dispersive spectral shape
% An additional feature separates the Bloch gain furthermore, as can be seen from the complex susceptibility of the Bloch and harmonic oscillators (Fig.\ref{fig_intro}a and b), where the optical gain is proportional to the imaginary part of the susceptibility.
%Due to its non-classical nature, Bloch gain is characterized with a dispersive shape, in sharp contrast to the well known gain of the harmonic oscillator defined with a symmetric Lorentzian. 
Analogous observations of this ubiquitous phenomenon are reported in other physical systems as well, e.g. Josephson junctions~\cite{averin1985bloch}, Bose-Einstein condensates~\cite{bendahan1996bloch}, complex potentials with PT symmetry~\cite{longhi2009bloch} and in optical~\cite{pertsch1999optical} and acoustic~\cite{sanchisAlepuz2007acoustic} waves.

%2. paragraph move to bloch gain in QCLs, e.g. start with "In QCLs, bloch gain was .." (refer to fig 1c)
More recently, pure quantum-mechanical treatments of the Bloch gain were developed in the density matrix formalism~\cite{willenberg2003intersubband} and the Green's function formalism~\cite{wacker2002gain}. They generalized the concept of the Bloch gain and showed that it is not exclusive to superlattices, but appears also between any two states (subbands) %between subbands% 
in semiconductor heterostructures, such as quantum cascade lasers (QCLs). QCLs are unipolar laser sources~\cite{faist1994quantum}, which emit in the mid-infrared~\cite{yao2012midIR} and terahertz~\cite{williams2007terahertz} spectral regions by nanoscale engineering of the conduction-band profile (Fig.~\ref{fig_intro}c).
%more on the following explain more detailed
%explain, linewidth broadening (dominantly due to ir scattering). jirauschek2017density 
The gain bandwidth of QCLs is broadened by elastic scattering processes beyond its natural limit defined with the carrier lifetimes ~\cite{ando1985linewidth,jirauschek2017density}. 
%Elastic scattering of carriers such as the interface roughness scattering greatly affects the operation of QCLs by broadening the linewidth beyond its natural limit defined by the carrier lifetimes~\cite{jirauschek2017density}.
%The transition linewidth of QCLs is broadened beyond its natural limit, given by the carrier lifetimes, dominantly due to elastic scattering processes such as interface roughness scattering~\cite{jirauschek2017density}. 
An accompanying effect of these processes, neglected by most researchers so far, is the occurrence of scattering-assisted optical transitions between subbands~\cite{willenberg2003intersubband,terazzi2007bloch}. %which connect states with different wavevectors (Fig.\ref{fig_intro}c).
%Additional crucial effect of elastic scattering is the occurrence of scattering-assisted optical transitions that spawn Bloch gain~\cite{terazzi2007bloch}, a key element needed for a correct model that has been so far neglected by most researchers.
%The inclusion of these second-order processes between a pair of subbands allows optical transitions that connect states with different wavevectors (Fig.\ref{fig_intro}c). 
They connect electronic states with nonidentical wavevectors and give rise to dispersive Bloch gain. The total gain is comprised of the Bloch contribution and the usual Lorentzian gain generated by the harmonic oscillator (Fig.~\ref{fig_intro}c).
%comprises this contribution and the usual symmetric Lorentzian gain (Fig.\ref{fig_intro}c).

%They connect states with nonidentical wavevectors and give rise to dispersive Bloch gain along with the usual symmetric Lorentzian gain (Fig.\ref{fig_intro}c).
%giving rise to optical gain as a sum of  contributions from the usual symmetric Lorentzian gain and dispersive Bloch gain which is independent on population inversion ~\cite{terazzi2007bloch}.

% By including the scattering-assisted optical transitions between a pair of subbands one allows transitions that are not momentum preserving (Fig.\ref{fig_intro}c). This results in optical gain that does not depend entirely on direct population inversionand can be divided into a contribution from a dispersive Bloch gain and the usual symmetric Lorentzian part(Fig.\ref{fig_intro}c). Consequently, the gain peak is red-shifted  with an overall asymmetric gainshape which has a paramount influence on the laser cavity dynamics, as will be shown.

In this work, we conduct a rigorous theoretical and numerical study of the Bloch gain and its influence on the laser dynamics. A meticulous simulation tool is developed which models and self-consistently couples every aspect of QCL operation -- from electronic band structure and charge transport to the light spatio-temporal evolution within the laser cavity.
We show that a dominant Bloch gain contribution is present in any operating QCL %at an elevated temperature 
and causes a giant Kerr nonlinearity at the laser wavelength. The induced nonlinearity plays an essential role in the laser cavity dynamics as it is a requirement for self-starting optical frequency combs~\cite{opacak2019theory}. Bloch gain is not only the reason why frequency modulated (FM) comb formation is predominantly found in dispersion compensated cavities~\cite{bidaux2017plasmon}, but it also allows tuning the laser into the phase turbulence regime. This can trigger the generation of soliton-like structures~\cite{meng2020midIR,piccardo2020freqeuncy}, establishing a bridge between semiconductor QCL lasers and Kerr microresonators~\cite{kippenberg2011microresonator}.

The spectral response of the laser active region 
is fully captured by its complex susceptibility $\chi = \chi_R + i \chi_I$. The optical gain is defined as $g=\omega\chi_I/n_rc$ with $\omega$ being the frequency, $n_r$ the refractive index and $c$ the speed of light.
The susceptibility that arises from %the interaction of 
any two subbands $u$ and $l$ in a semiconductor heterostructure is calculated as~\cite{willenberg2003intersubband}:
\begin{flalign}
\begin{split}
	&\chi(\omega)=\frac{\mu_{ul}^2\omega_0^2 }{\varepsilon_0  \omega^2 }\sum_{k}  \Big[ \frac{f_u(k)-f_l(k) }{(\hbar \omega-\mathsmaller{ \Delta} W(k))-i\gamma(k)} +  \\
	 &\highlight{i \frac{\gamma_u(k)(f_u(k_-)-f_u(k)) -  \gamma_l(k) (f_l(k_+)-f_l(k))} { \big(\hbar \omega-\mathsmaller{ \Delta} W(k)\big) \big( (\hbar \omega-\mathsmaller{ \Delta} W(k))  -i\gamma(k)  \big)  } }  \Big].
	\label{eq:1}
\end{split}
\end{flalign}
The total susceptibility in Eq. (\ref{eq:1}) comprises two components. The usual harmonic contribution is given by the first term in the square brackets in Eq. (\ref{eq:1}). It depends on the population inversion $f_u(k) - f_l(k)$ and yields a Lorentzian gainshape.  
%The usual population-inversion dependent harmonic component, which yields a Lorentzian gainshape, 
The dipole matrix element is $\mu_{ul}$, $\varepsilon_0$ is the vacuum permittivity, $f(k)$ and $\gamma(k)$ are the electron distribution and broadening at wavevector $k$ and $\mathsmaller{ \Delta} W(k) = W_u(k)-W_l(k)$ is the resonant transition energy, where $\mathsmaller{ \Delta} W_0=\mathsmaller{ \Delta} W(k=0) = \hbar \omega_0$. The highlighted second term in Eq. (\ref{eq:1}) is more intriguing. It generates the Bloch gain by allowing optical transitions between states with different wavevectors. Introduced notations are level broadenings $\gamma_{u,l}$, where $\gamma = \gamma_u + \gamma_l$ \cite{jirauschek2017density}, 
and in-plane momenta of the final states, defined as $k_{\pm}^2 = {\frac{m_{l,u}}{m_{u,l}}k^2 \pm \frac{2m_{l,u}}{\hbar^2}(\mathsmaller{ \Delta} W_0-\hbar \omega)}$. A thorough analysis is given in the Supplementary section 1.

%where $\mu_{ul}$ is the dipole matrix element and $\varepsilon_0$ is the vacuum permittivity.
%The electron distribution function of the subband $i$ at wavevector $k$ is $f_i(k)$ with the corresponding broadening $\gamma_i(k)$, where $\gamma = \gamma_u + \gamma_l$. The $k$-space resolved resonant transition energy is defined as $\mathsmaller{ \Delta} W(k) = W_u(k)-W_l(k)$, with $\mathsmaller{ \Delta} W_0=\mathsmaller{ \Delta} W(k=0) = \hbar \omega_0$ being the separation of the subband edges. Lastly, the in-plane momenta of the final states are defined with $k_{+,-}^2 = {\frac{m_{l,u}}{m_{u,l}}k^2 + \frac{2m_{l,u}}{\hbar^2}(\mathsmaller{ \Delta} W_0-\hbar \omega)}$, as can be seen in Fig.\ref{fig_intro}c. A thorough analysis and numerical implementation is given in the \textcolor{red}{supplementary material}.
%The total susceptibility in Eq. (\ref{eq:1}) comprises two contributions. The customary population-inversion dependent harmonic component that yields a Lorentzian gainshape is given with the first term in the square brackets in Eq. (\ref{eq:1}). %The second is the Bloch contribution that arises due to scattering assisted transitions which do not preserve the wavevector $k$, represented with the highlighted second term in the square brackets.
%Far more alluring is the highlighted second term which spawns the dispersive-shaped Bloch gain by allowing optical transitions between states with disparate wavevectors.

\begin{figure*}[ht]
	\centering
	\includegraphics[width = 1\textwidth]{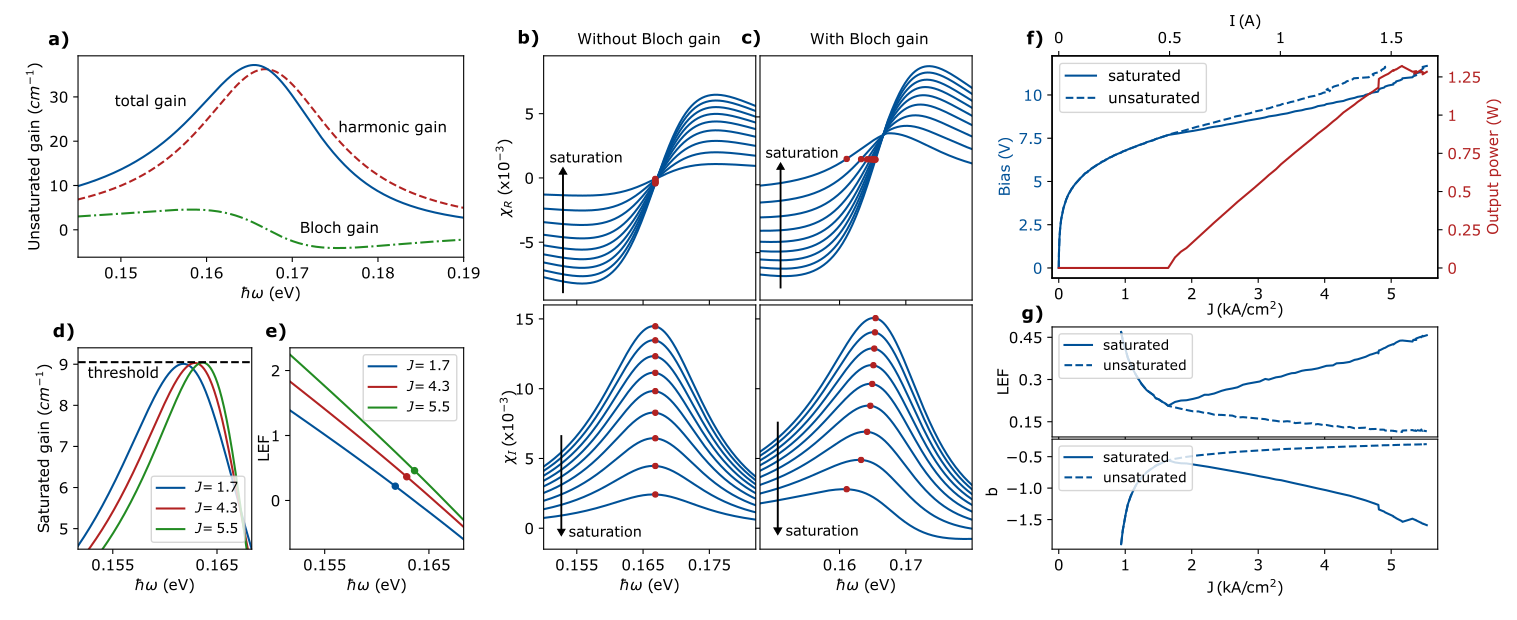}
	\caption{ \textbf{Bloch gain in QCLs.} \textbf{(a)} Unsaturated optical gain from Eq. (\ref{eq:1}) represented as the sum of the harmonic Lorentzian contribution and dispersive Bloch gain. \textbf{(b)},\textbf{(c)} Effect of saturation on real $\chi_R$ and imaginary part $\chi_I$ of the optical susceptibility, whether the Bloch gain is included or not. Red dots indicate values at gain peak and arrows show the direction of the increasing intensity. Bloch gain induces asymmetric gain, redshift and non-zero values of $\chi_R$ at the gain peak. \textbf{(d)} Saturated gain clamped to the threshold for three values of the current density $J$ $(\mathrm{kA/cm^2})$. \textbf{(e)} Spectrally calculated linewidth enhancement factor (LEF) for the same current densities as in (d). Values at the gain peak are indicated with dots. \textbf{(f)} Light-current-voltage (LIV) characteristic of the laser operating in continuous-wave at room temperature. \textbf{(g)} Dependence of the LEF and factor $b$ from Eq. (\ref{eq:2}) on the current density. A linear dependence of both values is observed.}
	\label{figGain}
\end{figure*}

While Eq. (\ref{eq:1}) provides an exact treatment of the Bloch gain, the origin of the dispersive spectral shape is not well understood.
%While Eq. (\ref{eq:1}) provides an exact treatment of the Bloch gain, the comprehension of the underlying mechanism which is responsible for the dispersive spectral shape remains elusive. 
It is not rare in physics to opt for a simpler model that provides an intuitive understanding over the exact one.
Bearing this in mind, Eq. (\ref{eq:1}) significantly reduces its complexity by assuming subband electron distributions in the frame of Boltzmann statistics. The electron concentration is usually low enough so that the Fermi-Dirac distribution reduces to the Boltzmann distribution and the carrier-carrier interaction is sufficiently large to enforce carrier thermalization~\cite{iotti2001carrier}.
%, justifying the approximation.
%This approximation is well based, since the electron concentration in high quality devices is low enough and  the carrier-carrier interaction is sufficiently large to enforce carrier thermalization~\cite{iotti2001carrier}.
Following the derivation presented in the Supplementary section 1.1, we analytically obtain a simplified definition of $\chi$:
\begin{align}
\begin{split}
\chi(\omega) &= \frac{\mu_{ul}^2 \omega_0^2}{\varepsilon_0  L_p \omega^2 }  \frac{ (n_u-n_l) + \highlight{i\mathsmaller{\frac{\gamma}{2k_BT}}(n_u+n_l)}   }{\hbar \omega-\mathsmaller{ \Delta} W_0 -i\gamma}  \\
&= \frac{\mu_{ul}^2 \omega_0^2}{\varepsilon_0  L_p \omega^2 }  \frac{n_u -n_l}{\hbar \omega-\mathsmaller{ \Delta} W_0 -i\gamma} \big( 1 + \highlight{ib}   \big),
\label{eq:2}
\end{split}
\end{align}
where $L_p$ is the QCL period length, $k_B$ is the Boltzmann constant, $T$ is the temperature and $n_{u,l}$ are the electron sheet densities of subbands ${u,l}$. Eq. (\ref{eq:2}) provides an understanding of the origin of Bloch gain, which is proportional to the highlighted terms. Contrary to the harmonic susceptibility, it is not dependent on the population inversion $(n_u-n_l)$ but rather on the population sum $(n_u+n_l)$.
The dispersive gainshape appears due to the imaginary value of the highlighted terms. They induce a $\pi/2$ phase shift and exchange the shapes of $\chi_R$ and $\chi_I$ (Fig.~\ref{fig_intro}a \& b). 
%They are independent of the population inversion $(n_u-n_l)$ and due to their imaginary value, a $\pi/2$ phase shift is introduced compared to the harmonic susceptibility, which results in the dispersive gainshape.
%Their robustness comes from the independence  on the population inversion $(n_u-n_l)$ and due to their imaginary value, the non-classical nature of the dispersive gainshape becomes evident, as a $\pi/2$ phase shift is introduced compared to the harmonic susceptibility.
%Due to their imaginary value, the non-classical nature of the scattering-assisted optical transitions and the dispersive gainshape they yield become evident as a $\pi/2$ phase shift is introduced in comparison to the harmonic susceptibility. The apparent independence of the Bloch gain on the direct population inversion $(n_u-n_l)$ proves its robustness and the necessity for its inclusion in a thorough study.
The factor $b$ in Eq. (\ref{eq:2}) captures the impact of the Bloch gain. It deviates the total gainshape from a Lorentzian curve and causes spectral asymmetry.
Subband nonparabolicity, which is known to induce similar behavior, has a weaker effect. 
Most importantly, Eq. (\ref{eq:2}) allows straightforward implementation of Bloch gain in any carrier transport model, unlike Eq. (\ref{eq:1}), which requires $k$-space resolved approaches.
%Bloch gain, determined with the factor $b$, has a key impact on the optical properties of the system because it deviates the gainshape from the symmetric Lorentzian curve defined with the expression in front of the parenthesis in Eq.(\ref{eq:2}).

%By considering the second-order scattering-assisted optical transitions, the standard susceptibility of a harmonic oscillator is modified by a complex multiplication term $(1+ib)$. The non-classical nature of the Bloch gain is captured with the imaginary value $ib$ as it introduces a $\pi/2$ phase shift compared to the harmonic susceptibility resulting in dispersive gainshape. Factor $b$ has a key impact on the optical properties of the system that govern the laser dynamics as it explains the deviation of the gainshape from the symmetric Lorentzian curve. It is proportional to the sum of the densities of two subbands $b\propto (n_u+n_l)$, making both it and the Bloch gain independent on the direct population inversion. 

With the aim of quantitatively assessing the influence of the Bloch gain on the laser dynamics, our model is employed to a reference QCL device~\cite{wittmann2008intersubband}. For details about our band structure and charge transport model, see the Methods section. 
%In this work, we apply our model to a reference QCL design~\cite{wittmann2008intersubband}. 
The calculated conduction-band profile with probability densities of the states and the electron density are shown in the Supplementary section 4.1. With the knowledge of the electron population, the calculation of the optical gain for the lasing transition follows from Eq. (\ref{eq:1}). Its unsaturated value is shown in Fig.~\ref{figGain}a. 
%since other subbands have a minor influence (\textcolor{red}{supplementary}). 
The Bloch gain induces an asymmetric total gainshape and a redshift of the peak. 
However, the unsaturated gain asymmetry conveys only a fraction of what happens above the laser threshold. 
%However, the unsaturated gain conveys only a fraction of the partial picture, as in an operating device, the lasing transition has a major influence on the electron distribution by depleting the direct population inversion.
%The light intensity will increase until the gain consequently saturates and clamps to equal the cavity loss.
In the usual harmonic description, the emission of light depletes the population inversion until the gain saturates to the threshold value, while $\chi_R$ remains zero at the gain peak (Fig.~\ref{figGain}b).
%The light intensity increases and consequently decreases the population inversion until the gain saturates and clamps to the cavity loss.
%In the usual harmonic description, this is reflected by a reduction of the available gain as the light intensity increases (Fig.\ref{figGain}c). 
On the other hand, the Bloch gain is independent of the population inversion and thus remains mostly unaffected. %affected much less by the emitted light.
As the harmonic gain fades away with stronger light intensity, the Bloch contribution prevails and results in an increasingly asymmetrical gain accompanied with a red-shift and non-zero $\chi_R$ at the gain peak (Fig.~\ref{figGain}c). Intriguingly, a negative global population inversion is required to completely diminish the total gain (Supplementary section 4.2). %, opening the possibility for lasing without population inversion~\cite{harris1989lasers}. 
%The lasing threshold is determined by total losses. Knowledge of its value makes the calculation of the exact saturated gain viable, as is shown in Fig.\ref{figGain}e for three different current densities $J$ through the device.
Fig.~\ref{figGain}d shows the saturated gain for three different current densities $J$.
%The gain becomes increasingly asymmetric towards the red end of the spectrum due to the increasing intensity in the cavity.
The gain peak is blueshifted due to the quantum-confined Stark effect and, more importantly, the asymmetry increases towards the dispersive shape. 
%The gain peak blue-shifts with the increasing bias due to the steepening of the conduction band profile and relocation of the eigenstates shown in Fig.\ref{figGain}a, which is known as the quantum-confined Stark effect.
%It is important to note, that the asymmetry still increases towards the red side of the spectrum.
%Additionally, the asymmetry increases towards a dispersive lineshape.
%Additionally, the gainshape deviates strongly from the symmetric Lorentzian curve, due to the increasing intensity of the emitted light. 

\begin{figure}[h!]
	\centering
	\includegraphics[width = 1\columnwidth]{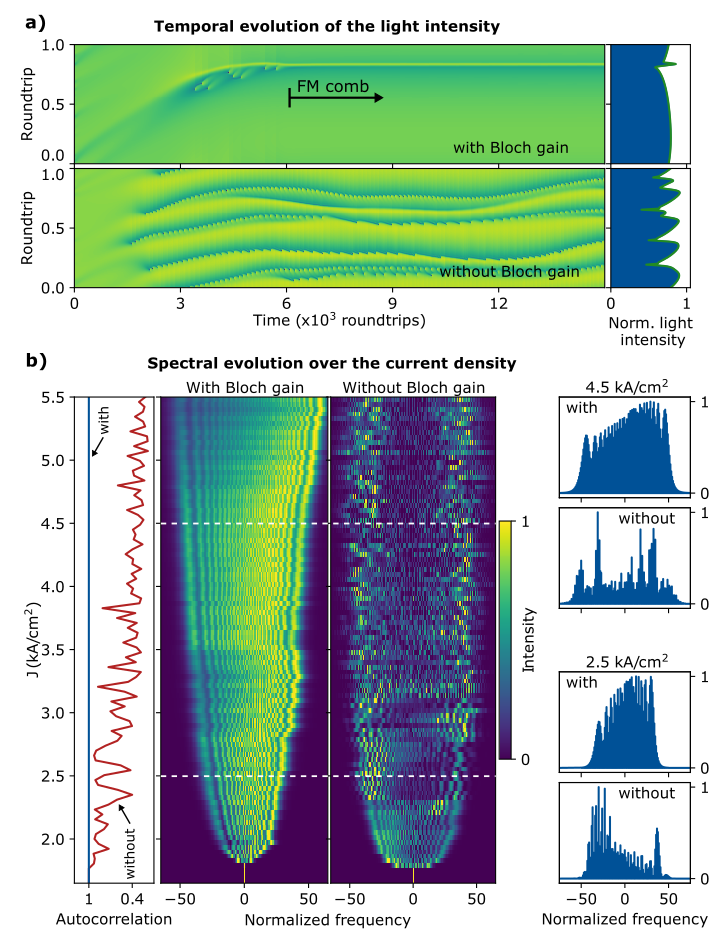}
	\caption{ \textbf{Frequency modulated (FM) combs in a Fabry-P\'{e}rot cavity.} \textbf{(a)} Temporal evolution of the light intensity depending on whether the Bloch gain is considered or not. The waveforms in the last roundtrip are shown on the right. The consideration of Bloch gain leads to an FM comb formation after 6000 roundtrips. Excluding it results in a unlocked evolution of the field. \textbf{(b)} Evolution of the intensity spectrum with the increasing current density $J$. Near threshold, the spectrum consists of a single mode and broadens with the increase of $J$. The Bloch gain induced Kerr nonlinearity forms a locked FM comb over the entire range of $J$, as is seen from the autocorrelation value equal to unity (blue line on the left). Neglecting the Bloch contribution leads to unlocked states with the autocorrelation value smaller than unity (red line on the left). The four spectra on the right are taken at $2.5~\mathrm{kA/cm^2}$ and $4.5~\mathrm{kA/cm^2}$, indicated by the white dashed lines.}
	\label{figCavity_FM}
\end{figure}

%We identified Bloch gain to be the dominant contribution to the gain asymmetry in mid-infrared QCLs at elevated temperatures. Smaller contributions can be due to non-parabolicity or off-resonant transition, where the latter should be considered in THz QCLs. 

Gain asymmetry has historically been analyzed in the context of the laser linewidth broadening.
%Gain asymmetry has historically been analyzed mostly in interband lasers due to its impact on the Schawlow-Townes-Limit, which describes the quantum noise limited linewidth~\cite{}. 
It was treated with the empirical linewidth enhancement factor (LEF)~\cite{Henry1982Theory}, defined as $\mathrm{LEF}=-(\nicefrac{\partial \chi_R}{\partial N}) / (\nicefrac{\partial \chi_I}{\partial N})$, where $N$ is the carrier population.
In interband lasers, the gain asymmetry and LEF dominantly originate from the opposite curvature of the valence and conduction band.
%the inherent opposite curvature of the valence and conduction band.
%inherent bending of the valence and conduction bands in opposite directions.
%It was empirically treated with the so called linewidth enhancement factor (LEF)~\cite{Henry1982Theory}, defined as $\mathrm{LEF}=-(\nicefrac{\partial \chi_R}{\partial N}) / (\nicefrac{\partial \chi_I}{\partial N})$, where $N$ is the number of carriers.
Since both laser levels in a QCL have similar curvatures, the LEF was expected to vanish. Interestingly, non-zero experimental values were obtained mostly between -0.5 and 1.5~\cite{green2008linewidth,jumpertz2016measurements}. 
We explain this with the gain asymmetry in QCLs that is dominantly caused by the Bloch gain. Furthermore, the LEF is frequency dependent, which yields a large range of values shown in Fig.~\ref{figGain}e. 
Elimination of the Bloch gain in QCLs yields a symmetric gain profile and vanishing LEF (Supplementary secion 4.2). 
%The LEF in QCLs dominantly stems from the Bloch gain, as its absence removes the gain asymmetry and leads to vanishing values (Supplementary Material).
%The large range of reported values is explained through a strong frequency dependence of the LEF, as the measurement frequency is detuned from the gain peak position. 
%This is seen in Fig.~\ref{figGain}f, where the values at the gain peak are indicated with dots. 
%Conversely, LEF value remains zero by omitting the Bloch gain (Supplementary Material).
%Rather than relying on a description with the LEF, which is frequency dependent, we incorporate the Bloch gain induced asymmetry with the factor $b$, which is naturally imposed in Eq. (\ref{eq:2}).
Fig.~\ref{figGain}f shows the simulated light-current-voltage (LIV) characteristic with the lasing threshold at around $J=1.6~\mathrm{kA/cm^2}$ and rollover at $J=5.5~\mathrm{kA/cm^2}$ \cite{wittmann2008intersubband}. %, as a validation of the carrier transport model.
The calculated values of the LEF and factor $b$ at the gain peak for the entire range of the current density $J$ from the LIV are shown in Fig.~\ref{figGain}g. 
%Values of both LEF and factor $b$ have been calculated for the entire dynamic range of the current $J$ from the LIV and shown in Fig.\ref{figGain}h.
Although the population inversion is clamped, the population sum increases with the current in Eq. (\ref{eq:2}). This leads to a linear dependence of the LEF and factor $b$ on $J$,
which matches observations found in literature~\cite{green2008linewidth,jumpertz2016measurements,piccardo2019frequency}. % despite the vast range of reported values.
%Conversely, the exclusion of Bloch gain leads to vanishing LEF, as can be seen in the Supplementary Material.
%A clear linear dependence is observed, matching the literature~\cite{jumpertz2016measurements}. 
The impact of the gain saturation is underlined yet again, as the saturated values notably break off from the unsaturated ones. % above the lasing threshold.

The gain asymmetry causes changes of both the gain and the refractive index of the active region. 
%Physical consequences of the Bloch gain induced asymmetry are reflected in the alterations of both the gain and the refractive index in the laser active region. 
Their alterations are induced by variable electron population, as was described by Agrawal~\cite{Agrawal1988Population}.
%mention Agrawal -> modulation of the real part of refractive index
%maybe move definition of lef here
%In contrast to interband lasers, 
%QCL ps lifetime
%Giant Kerr nonlinearity    (main message of this paragraph)
This is closely related to a dependence of the gain and refractive index on the intensity (Fig.~\ref{figGain}c), which gives rise to a Kerr nonlinearity.
%This closely relates to the light intensity imposed variations of the gain and refractive index (Fig.\ref{figGain}d), which give rise to a Kerr nonlinearity. 
Although the bulk nonlinearity of a semiconductor crystal is small, the resonant contribution from the asymmetric nature of the gain yields a giant Kerr nonlinearity due to ultrafast dynamics in QCLs ~\cite{opacak2019theory}. 
%Although the bulk nonlinearity of a semiconductor crystal is small, its resonant contribution stemming from the asymmetric nature of the gain results in a giant Kerr nonlinearity in QCLs due to their ultrafast dynamics~\cite{opacak2019theory}. 
Based on the saturation analysis of $\chi$ in Supplementary section 3, we calculate the resonant Kerr contribution due to Bloch gain to be in the range of $\mathrm{10^{-15}~m^2/W}$, which is two orders of magnitude larger than the highest bulk values~\cite{gaeta2019photonic}. 
%Conversely, in interband lasers which have a stronger gain asymmetry, the resonant Kerr contribution would be minuscule due to slower gain dynamics.

%start intro n
%Optical nonlinearities couple the amplitude and the phase of the laser field -> frequency comb
Optical nonlinearities couple the amplitude and the phase of the intracavity laser field and give rise to coherent processes such as frequency comb formation. Frequency combs are lasers whose spectra consist of equidistant modes with a fixed phase relation~\cite{haensch2006nobel,hall2006nobel}. Although historically their formation relied on the emission of short pulses~\cite{Keilmann2004Time}, recently a new type of frequency modulated (FM) combs is blossoming. They are self-starting and appear in numerous Fabry-P\'{e}rot laser types such as QCLs~\cite{hugi2012mid}, interband cascade lasers~\cite{schwarz2019monolithic}, quantum dot lasers~\cite{hillbrand2020inphase} and laser diodes~\cite{sterczewski2020frequency}. The fascinating property of FM combs, which distinguishes them from other frequency combs, is an almost constant intensity accompanied with a linear frequency chirp~\cite{singleton2018evidence}.
%The fascinating property of FM combs that distinguishes them is an almost constant intensity accompanied with a linear frequency chirp~\cite{singleton2018evidence}. 
This unique behavior was explained in~\cite{opacak2019theory} as a result of the group velocity dispersion (GVD) or more importantly a Kerr nonlinearity, thus bringing the role of the Bloch gain in FM comb formation to the foreground.
%

%Lastly, we will analyze the influence of the Bloch gain on the laser field dynamics and its role in the optical frequency comb formation. Frequency combs are lasers whose spectra consist of equidistant modes with a fixed phase relation~\cite{haensch2006nobel,hall2006nobel}. Although historically their formation relied on the emission of short pulses~\cite{Keilmann2004Time}, recently a new type of frequency modulated (FM) combs are blossoming due to their self-starting nature and  sponataneous appearance in numerous laser types such as QCLs~\cite{hugi2012mid}, interband cascade lasers~\cite{schwarz2019monolithic}, quantum dot lasers~\cite{hillbrand2020inphase} and laser diodes~\cite{sterczewski2020frequency}. The fascinating property of FM combs that distinguishes them is an almost constant laser intensity accompanied with a linear frequency chirp. This unique behavior was first explained in~\cite{opacak2019theory} as a result of the group velocity dispersion (GVD) or Kerr nonlinearity, which is the case in this work. Although the bulk nonlinearity of a semiconductor crystal is small, its resonant contributions stemming from the asymmetric nature of the gain spectra gives rise to a giant Kerr nonlinearity~\cite{opacak2019theory} in QCLs. The latter describes the light intensity induced refractive index changes, thus coupling the amplitude and the phase of the laser field which governs frequency comb formation.

%
In order to quantitatively study the frequency comb dynamics, we conduct spatio-temporal simulations of the intracavity field based on a master equation approach~\cite{opacak2019theory}. 
We describe the gain shape asymmetry through the parameter $b$ from Eq. (\ref{eq:2}). Its population dependence can be accurately modeled as a function of the current density $J$ and the laser field intensity $I$ (Supplementary equation (42)).
This allows a self-consistent implementation into the master equation to include the Bloch gain:
\begin{align}
\begin{split}
\Big( \frac{n}{c}&\partial_t  \pm \partial_z \Big) E_\pm = \frac{g}{2} \frac{1+ib}{1+i\xi} \Big[ E_\pm - \tilde{T}_2 \partial_t E_\pm + \tilde{T}_2^2 \partial_t^2 E_\pm \Big]\\
&-   \frac{gT_g}{T_1 I_\mathrm{sat}}  \frac{1+ib}{1+i\xi}\Big[ \abs{E_\mp}^2 E_\pm - ( \tilde{T}_2+T_\mathrm{g}) \abs{E_\mp}^2 \partial_t E_\pm \\
&-(\tilde{T}_2+T_\mathrm{I}) E_\pm E_\mp \partial_t E^*_\mp - \tilde{T}_2 E_\pm E^*_\mp \partial_t E_\mp  \Big] -\frac{\alpha_w}{2}  E_{\pm},
\label{eq:3} 
\end{split}
\end{align}
where $E\pm$ are the right and left propagating field envelopes, $T_1,T_2$ and $T_g$ are the recovery times of the gain, polarization and the population grating, $\alpha_w$ is the waveguide loss, $g$ is the saturated gain, $I_\mathrm{sat}$ is the saturation intensity and $I = \abs{E_+}^2\!+\!\abs{E_-}^2$ the normalized intensity. The Bloch gain enters the equation through terms $b$, $\xi(b)$ and $\tilde{T}_2(b)$. A detailed derivation is presented in the Supplementary section 2.3, along with the analysis for interband lasers with slower dynamics. 

The numerical results for a Fabry-P\'{e}rot QCL are shown in Fig.~\ref{figCavity_FM}. Using Eq. (\ref{eq:3}), we simulate 30 000 roundtrips of the electric field evolution to ensure that a steady state has been reached. %Such long simulation times are possible due to a highly efficient numeric implementation that runs in parallel on 4000 GPU threads using the CUDA library~\cite{cuda}. 
Temporal evolution of the light intensity for one bias point is shown in Fig.~\ref{figCavity_FM}a. The inclusion of the Bloch gain leads to a periodic waveform after 6000 roundtrips and the formation of an FM comb, which is fully characterized in the Supplementary section 4.2. Conversely, the intensity evolves chaotically in the absence of a locking mechanism provided by the Bloch gain induced Kerr nonlinearity.
By extracting the scattering rates from the transport model, we are able to accurately simulate the intracavity dynamics from the laser threshold to rollover (Fig.~\ref{figCavity_FM}b). 
The laser is in the single mode regime near the threshold and significantly broadens its spectrum with the current increase.
The key role of the Bloch gain is clearly visible, as it leads to an FM comb operation over the entire bias range. This is indicated by the autocorrelation value equal to one in Fig.~\ref{figCavity_FM}b. 
In sharp contrast, the pure harmonic gain results in unlocked states with the autocorrelation smaller than unity and chaotic spectra.
This validates the Bloch gain induced giant Kerr nonlinearity as an efficient locking mechanism and explains why FM combs in QCLs have mostly been found in GVD compensated cavities~\cite{bidaux2017plasmon,singleton2018evidence,hillbrand2018tunable}. The interplay with a non-zero GVD yields an unlocked state for most of the bias range (Supplementary section 4.4), in accordance with literature~\cite{hillbrand2018tunable}. 

\begin{figure}[t]
	\centering
	\includegraphics[width = 1\columnwidth]{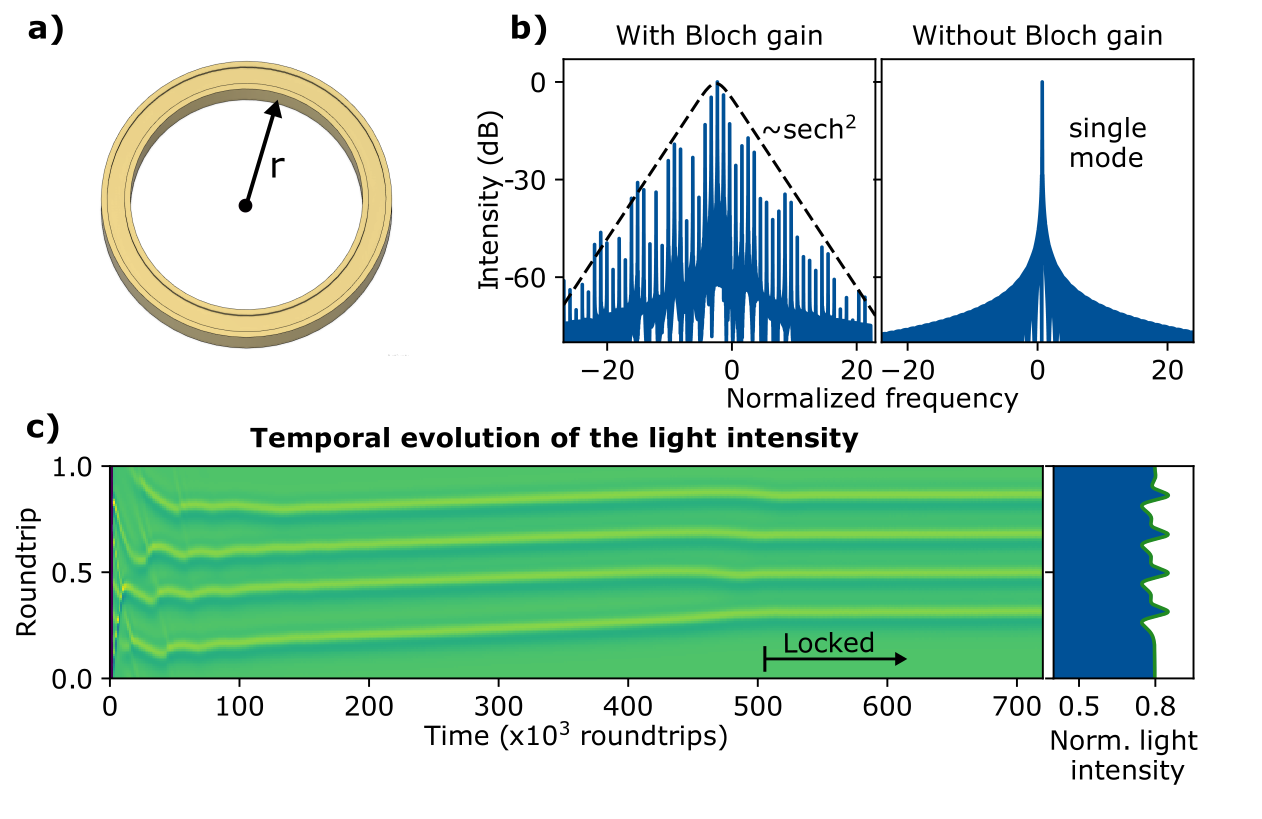}
	\caption{ \textbf{Spatial patterns in a monolithic ring laser comb. } \textbf{(a)} Schematic of a ring cavity laser. \textbf{(b)} Intensity spectra of a ring QCL. Bloch gain leads to a multi-mode instability through phase turbulence~\cite{piccardo2020freqeuncy}. A $\mathrm{sech^2}$ envelope is fitted to the spectrum. Elimination of Bloch gain yields single-mode operation. \textbf{(c)} Temporal evolution of the intensity shows an initial turbulent regime that forms a frequency comb after 510~000 roundtrips.  }
	\label{figCavity_ring}
\end{figure}

Linking the physics of FM combs to Bloch gain induced giant Kerr nonlinearity suggests a connection to the Kerr combs in microresonators~\cite{kippenberg2011microresonator}. They represent passive media, where pumping is achieved through an external injection of a monochromatic laser and the gain stems from the Kerr nonlinearity of the bulk crystal. Through a cascaded parametric process, the injected wave induces the appearance of side-modes giving rise to phase-locked frequency combs in the form of temporal solitons~\cite{herr2013temporal}.
%In the latter, an injected monochromatic field is destabilized by the Kerr nonlinearity of the bulk crystal giving rise to phase-locked frequency combs in the form of temporal solitons~\cite{herr2013temporal}. 
%Although they rely on fundamentally different governing mechanisms, 
QCL combs in ring cavities (Fig.~\ref{figCavity_ring}a) have recently been shown to possess several similarities with Kerr microresonators~\cite{piccardo2019frequency,meng2020midIR}. %despite the fact that they are electrically pumped.
%Although they rely on fundamentally different governing mechanisms, ring QCL combs have recently been connected with the formation of dissipative Kerr solitons~\cite{piccardo2019frequency}. 
Within the framework of the Ginzburg-Landau formalism~\cite{aranson2002ginzburg}, it was demonstrated that a single mode operation is destabilized by the phase turbulence. The latter is controlled with the laser nonlinearity to induce multi-mode emission with a sech-type spectrum (Fig.~\ref{figCavity_ring}b). 
This can trigger the formation of localized structures in the waveform (Fig.~\ref{figCavity_ring}c), which are related to dissipative Kerr solitons. In the absence of the Bloch gain, the laser operates in a single-mode regime with constant intensity. The study in~\cite{columbo2020unifying} demonstrated that the Kerr microresonators and ring QCLs can both be analyzed within the same theoretical framework and predicted the emission of temporal solitons from a ring QCL with a suitable nonlinearity.
%Know we know Bloch gain is responsible to trigger the phase turbulence regime. It can not only be simulated self consistently, it explains how to specifically optimize QCLs for solitone like structures with wider spectral bandwidth
%As we now know that the Kerr nonlinearity dominantly stems from the Bloch gain, we could realize new methods of QCL comb formation with wider spectral bandwidth via optimization of the gainshape by using our model to carefully optimize the gainshape for soliton emission, we could realize new methods of QCL comb formation with wider spectral bandwidth.
As we now know that the Kerr nonlinearity dominantly stems from the Bloch gain, by using our model to carefully optimize the gainshape for soliton emission, new methods of QCL comb formation with wider spectral bandwidth could be realized.

%main message:
%we identified Bloch gain as the main mechanism that triggers comb formation in quantum cascade lasers. Bloch gain is due to .. .... which is general to intersubband gain at elevated temperatures and thus present in virtually all mid-infrared QCLs. 

%Bloch gain 
%A self-consistent simulation model was built to study QCL operation altogether. It showed that the saturated gain deviates from a Lorentzian curve towards a dispersive asymmetric shape accompanied with a peak redshift due to Bloch gain. We connect the gain asymmetry to the LEF and explain its experimentally obtained values and linear dependence on the current. 
%Freuqnecy comb generation in Fabry-P\'{e}rot lasers by Bloch gain can operate over the entire multi-mode bias range under optimal condition.
%different in other semiconductor lasers, where dispersion need to be precisely adjusted to match the optimal condition for a particular laser bias.

%Finally, the .... of Bloch gain and dispersion, enables 
%which is very different compared to dispersion triggered comb generation. %maybe This is very different to dispersion triggered comb generation, which is restricted to a narrow range

In conclusion, we have shown that a substantial Bloch gain contribution is present in QCLs due to scattering-assisted optical transitions and that it provides a coherent locking mechanism for frequency comb formation.
A self-consistent simulation model was built to study QCL operation altogether, including bands tructure, carrier transport and cavity field dynamics. It showed that the saturated gain considerably deviates from a Lorentzian curve towards a dispersive asymmetric shape due to Bloch gain. We connect the gain asymmetry to the LEF and explain its experimentally obtained values. % and linear dependence on the current. 
The nonclassical nature of the Bloch gain is captured with a single population-dependent parameter $b$.  %that depends on the laser current and light intensity.
It is implemented in the master equation %that includes the Bloch gain self-consistently
to study the spatio-temporal evolution of the laser field. In accordance with this, we discover that
FM comb formation in Fabry-P\'{e}rot QCLs is triggered by a Bloch gain induced giant Kerr nonlinearity.
%It is induced by the Bloch gain in QCLs due to ultrafast dynamics  two orders of magnitude larger compared to record bulk values . %Due to their ultra-fast dynamics, the induced Kerr nonlinearity is two orders of magnitude larger than in bulk material.  
%Due to the ultra-fast gain dynamics of QCLs, gain asymmetry% leads to carrier induced refractive index and gain modulations that 
%spawns a giant Kerr nonlinearity that triggers FM comb formation.
The Bloch gain therefore acts as a efficient locking mechanism for the entire range of the bias current, which explains why FM combs were experimentally observed mostly in GVD compensated cavities.
In a ring resonator, the impact of Bloch gain is particularly strong, due to the low cavity losses and the stronger saturation.
The induced Kerr nonlinearity destabilizes the single mode operation through phase turbulence and can result in comb formation and the emission of localized structures. This paves the way towards broadband active Kerr combs in the mid-infrared range.
%which suggests the possibility of active Kerr combs.
%that serve as a link towards passive Kerr combs.
By careful design of the laser active region and cavity, the Bloch contribution to the total gain can be controlled in order to tailor the induced LEF and Kerr nonlinearity. This would allow us to further optimize QCL frequency combs for broadband emission and discover new states of light.

\vspace{-0.2cm}

\section*{Methods}

\vspace{0.1cm}

\setstretch{1.}
\noindent\textbf{Band structure and electron transport}: The band structure of the device is calculated using an envelope function formalism in the two-band $k \cdot p$ model~\cite{bastard1988wave}. Electron transport between the states is computed by employing a one-dimensional density matrix approach~\cite{jirauschek2017density}, which includes the energy-preserving tunneling~\cite{terazzi2008sequential,terazzi2010density}. The electron density is calculated self-consistently and accounts for the most relevant scattering processes via longitudinal optical phonons, acoustic phonons, interface roughness and alloy scattering~\cite{jirauschek2014modeling,harrison2003quantum}.

\vspace{0.2cm}

\noindent\textbf{Laser cavity model}: In order to study frequency comb dynamics, we use the master equation (\ref{eq:3}). We simulate 30~000 roundtrips of spatio-temporal laser field evolution in Fabry-P\'{e}rot cavities and 720~000 roundtrips in ring cavities to ensure that the laser is in a frequency comb regime. Such long simulation times are allowed by a highly efficient numeric implementation using the CUDA library~\cite{cuda}. The code is parallelized and runs on 2000 threads on the graphics processing unit (GPU) within a PC, which resulted in a  speed-up factor of 500 compared to the implementation on a central processing unit. We used an NVIDIA GeForce GTX 1070 Ti GPU. As an example, the spatio-temporal simulation shown in Fig.~\ref{figCavity_ring}c, which consists of 720 million time steps, took 32 minutes to run. The values of the parameters that are used in the cavity model are listed in the Supplementary section 5.

\footnotesize

\providecommand{\noopsort}[1]{}\providecommand{\singleletter}[1]{#1}

\bibliographystyle{naturemag_noURL}
%\bibliography{literature}
\bibliography{mybib}

\begin{thebibliography}{10}
\expandafter\ifx\csname url\endcsname\relax
  \def\url#1{\texttt{#1}}\fi
\expandafter\ifx\csname urlprefix\endcsname\relax\def\urlprefix{URL }\fi
\providecommand{\bibinfo}[2]{#2}
\providecommand{\eprint}[2][]{\url{#2}}

\bibitem{bloch1929ueber}
\bibinfo{author}{Bloch, F.}
\newblock \bibinfo{title}{{\"U}ber die {Q}uantenmechanik der {E}lektronen in
  {K}ristallgittern}.
\newblock \emph{\bibinfo{journal}{Zeitschrift f\"ur Physik}}
  \textbf{\bibinfo{volume}{52}}, \bibinfo{pages}{555--600}
  (\bibinfo{year}{1929}).

\bibitem{zener1934theory}
\bibinfo{author}{Zener, C.}
\newblock \bibinfo{title}{A theory of the electrical breakdown of solid
  dielectrics}.
\newblock \emph{\bibinfo{journal}{Proceedings of the Royal Society of London.
  Series A, Containing Papers of a Mathematical and Physical Character}}
  \textbf{\bibinfo{volume}{145}}, \bibinfo{pages}{523--529}
  (\bibinfo{year}{1934}).

\bibitem{waschke1993coherent}
\bibinfo{author}{Waschke, C.} \emph{et~al.}
\newblock \bibinfo{title}{Coherent submillimeter-wave emission from bloch
  oscillations in a semiconductor superlattice}.
\newblock \emph{\bibinfo{journal}{Physical Review Letters}}
  \textbf{\bibinfo{volume}{70}}, \bibinfo{pages}{3319--3322}
  (\bibinfo{year}{1993}).

\bibitem{ashcroft1976solid}
\bibinfo{author}{Ashcroft, N.~W.} \& \bibinfo{author}{Mermin, D.~N.}
\newblock \emph{\bibinfo{title}{Solid State Physics}}
  (\bibinfo{publisher}{Brooks Cole}, \bibinfo{year}{1976}).

\bibitem{esaki1970superlattice}
\bibinfo{author}{Esaki, L.} \& \bibinfo{author}{Tsu, R.}
\newblock \bibinfo{title}{Superlattice and negative differential conductivity
  in semiconductors}.
\newblock \emph{\bibinfo{journal}{{IBM} Journal of Research and Development}}
  \textbf{\bibinfo{volume}{14}}, \bibinfo{pages}{61--65}
  (\bibinfo{year}{1970}).

\bibitem{leo1992observation}
\bibinfo{author}{Leo, K.}, \bibinfo{author}{Bolivar, P.~H.},
  \bibinfo{author}{Br\"{u}ggemann, F.}, \bibinfo{author}{Schwedler, R.} \&
  \bibinfo{author}{K\"{o}hler, K.}
\newblock \bibinfo{title}{Observation of bloch oscillations in a semiconductor
  superlattice}.
\newblock \emph{\bibinfo{journal}{Solid State Communications}}
  \textbf{\bibinfo{volume}{84}}, \bibinfo{pages}{943--946}
  (\bibinfo{year}{1992}).

\bibitem{feldmann1992optical}
\bibinfo{author}{Feldmann, J.} \emph{et~al.}
\newblock \bibinfo{title}{Optical investigation of bloch oscillations in a
  semiconductor superlattice}.
\newblock \emph{\bibinfo{journal}{Physical Review B}}
  \textbf{\bibinfo{volume}{46}}, \bibinfo{pages}{7252--7255}
  (\bibinfo{year}{1992}).

\bibitem{ktitorov1971bragg}
\bibinfo{author}{Ktitorov, S.~A.}, \bibinfo{author}{Simin, G.~S.} \&
  \bibinfo{author}{Sindalovskii, V.~Y.}
\newblock \bibinfo{title}{Bragg reflections and the high-frequency conductivity
  of an electronic solid-state plasma}.
\newblock \emph{\bibinfo{journal}{Fiz. Tver. Tela}}
  \textbf{\bibinfo{volume}{13}}, \bibinfo{pages}{2230--2233}
  (\bibinfo{year}{1971}).

\bibitem{sekine2005dispersive}
\bibinfo{author}{Sekine, N.} \& \bibinfo{author}{Hirakawa, K.}
\newblock \bibinfo{title}{Dispersive terahertz gain of a nonclassical
  oscillator: Bloch oscillation in semiconductor superlattices}.
\newblock \emph{\bibinfo{journal}{Physical Review Letters}}
  \textbf{\bibinfo{volume}{94}} (\bibinfo{year}{2005}).

\bibitem{averin1985bloch}
\bibinfo{author}{Averin, D.~V.}, \bibinfo{author}{Zorin, A.~B.} \&
  \bibinfo{author}{Likharev, K.~K.}
\newblock \bibinfo{title}{Bloch oscillations in small josephson junctions}.
\newblock \emph{\bibinfo{journal}{Sov. Phys. JETP}}
  \textbf{\bibinfo{volume}{61}}, \bibinfo{pages}{407--413}
  (\bibinfo{year}{1985}).

\bibitem{bendahan1996bloch}
\bibinfo{author}{Dahan, M.~B.}, \bibinfo{author}{Peik, E.},
  \bibinfo{author}{Reichel, J.}, \bibinfo{author}{Castin, Y.} \&
  \bibinfo{author}{Salomon, C.}
\newblock \bibinfo{title}{Bloch oscillations of atoms in an optical potential}.
\newblock \emph{\bibinfo{journal}{Physical Review Letters}}
  \textbf{\bibinfo{volume}{76}}, \bibinfo{pages}{4508--4511}
  (\bibinfo{year}{1996}).

\bibitem{longhi2009bloch}
\bibinfo{author}{Longhi, S.}
\newblock \bibinfo{title}{Bloch oscillations in complex crystals
  {withPTSymmetry}}.
\newblock \emph{\bibinfo{journal}{Physical Review Letters}}
  \textbf{\bibinfo{volume}{103}} (\bibinfo{year}{2009}).

\bibitem{pertsch1999optical}
\bibinfo{author}{Pertsch, T.}, \bibinfo{author}{Dannberg, P.},
  \bibinfo{author}{Elflein, W.}, \bibinfo{author}{Br\"{a}uer, A.} \&
  \bibinfo{author}{Lederer, F.}
\newblock \bibinfo{title}{Optical bloch oscillations in temperature tuned
  waveguide arrays}.
\newblock \emph{\bibinfo{journal}{Physical Review Letters}}
  \textbf{\bibinfo{volume}{83}}, \bibinfo{pages}{4752--4755}
  (\bibinfo{year}{1999}).

\bibitem{sanchisAlepuz2007acoustic}
\bibinfo{author}{Sanchis-Alepuz, H.}, \bibinfo{author}{Kosevich, Y.~A.} \&
  \bibinfo{author}{S{\'{a}}nchez-Dehesa, J.}
\newblock \bibinfo{title}{Acoustic analogue of electronic bloch oscillations
  and resonant zener tunneling in ultrasonic superlattices}.
\newblock \emph{\bibinfo{journal}{Physical Review Letters}}
  \textbf{\bibinfo{volume}{98}} (\bibinfo{year}{2007}).

\bibitem{willenberg2003intersubband}
\bibinfo{author}{Willenberg, H.}, \bibinfo{author}{D\"{o}hler, G.~H.} \&
  \bibinfo{author}{Faist, J.}
\newblock \bibinfo{title}{Intersubband gain in a bloch oscillator and quantum
  cascade laser}.
\newblock \emph{\bibinfo{journal}{Physical Review B}}
  \textbf{\bibinfo{volume}{67}} (\bibinfo{year}{2003}).

\bibitem{wacker2002gain}
\bibinfo{author}{Wacker, A.}
\newblock \bibinfo{title}{Gain in quantum cascade lasers and superlattices: A
  quantum transport theory}.
\newblock \emph{\bibinfo{journal}{Physical Review B}}
  \textbf{\bibinfo{volume}{66}} (\bibinfo{year}{2002}).

\bibitem{faist1994quantum}
\bibinfo{author}{Faist, J.} \emph{et~al.}
\newblock \bibinfo{title}{Quantum cascade laser}.
\newblock \emph{\bibinfo{journal}{Science}} \textbf{\bibinfo{volume}{264}},
  \bibinfo{pages}{553} (\bibinfo{year}{1994}).

\bibitem{yao2012midIR}
\bibinfo{author}{Yao, Y.}, \bibinfo{author}{Hoffman, A.~J.} \&
  \bibinfo{author}{Gmachl, C.~F.}
\newblock \bibinfo{title}{Mid-infrared quantum cascade lasers}.
\newblock \emph{\bibinfo{journal}{Nature Photonics}}
  \textbf{\bibinfo{volume}{6}}, \bibinfo{pages}{432--439}
  (\bibinfo{year}{2012}).

\bibitem{williams2007terahertz}
\bibinfo{author}{Williams, B.~S.}
\newblock \bibinfo{title}{Terahertz quantum-cascade lasers}.
\newblock \emph{\bibinfo{journal}{Nature Photonics}}
  \textbf{\bibinfo{volume}{1}}, \bibinfo{pages}{517--525}
  (\bibinfo{year}{2007}).

\bibitem{ando1985linewidth}
\bibinfo{author}{Ando, T.}
\newblock \bibinfo{title}{Line width of inter-subband absorption in inversion
  layers: Scattering from charged ions}.
\newblock \emph{\bibinfo{journal}{Journal of the Physical Society of Japan}}
  \textbf{\bibinfo{volume}{54}}, \bibinfo{pages}{2671--2675}
  (\bibinfo{year}{1985}).

\bibitem{jirauschek2017density}
\bibinfo{author}{Jirauschek, C.}
\newblock \bibinfo{title}{Density matrix monte carlo modeling of quantum
  cascade lasers}.
\newblock \emph{\bibinfo{journal}{Journal of Applied Physics}}
  \textbf{\bibinfo{volume}{122}}, \bibinfo{pages}{133105}
  (\bibinfo{year}{2017}).

\bibitem{terazzi2007bloch}
\bibinfo{author}{Terazzi, R.} \emph{et~al.}
\newblock \bibinfo{title}{Bloch gain in quantum cascade lasers}.
\newblock \emph{\bibinfo{journal}{Nature Physics}}
  \textbf{\bibinfo{volume}{3}}, \bibinfo{pages}{329--333}
  (\bibinfo{year}{2007}).

\bibitem{opacak2019theory}
\bibinfo{author}{Opa{\v{c}}ak, N.} \& \bibinfo{author}{Schwarz, B.}
\newblock \bibinfo{title}{Theory of frequency-modulated combs in lasers with
  spatial hole burning, dispersion, and kerr nonlinearity}.
\newblock \emph{\bibinfo{journal}{Physical Review Letters}}
  \textbf{\bibinfo{volume}{123}} (\bibinfo{year}{2019}).

\bibitem{bidaux2017plasmon}
\bibinfo{author}{Bidaux, Y.} \emph{et~al.}
\newblock \bibinfo{title}{Plasmon-enhanced waveguide for dispersion
  compensation in mid-infrared quantum cascade laser frequency combs}.
\newblock \emph{\bibinfo{journal}{Optics Letters}}
  \textbf{\bibinfo{volume}{42}}, \bibinfo{pages}{1604} (\bibinfo{year}{2017}).

\bibitem{meng2020midIR}
\bibinfo{author}{Meng, B.} \emph{et~al.}
\newblock \bibinfo{title}{Mid-infrared frequency comb from a ring quantum
  cascade laser}.
\newblock \emph{\bibinfo{journal}{Optica}} \textbf{\bibinfo{volume}{7}},
  \bibinfo{pages}{162} (\bibinfo{year}{2020}).

\bibitem{piccardo2020freqeuncy}
\bibinfo{author}{Piccardo, M.} \emph{et~al.}
\newblock \bibinfo{title}{Frequency combs induced by phase turbulence}.
\newblock \emph{\bibinfo{journal}{Nature}} \textbf{\bibinfo{volume}{582}},
  \bibinfo{pages}{360--364} (\bibinfo{year}{2020}).

\bibitem{kippenberg2011microresonator}
\bibinfo{author}{Kippenberg, T.~J.}, \bibinfo{author}{Holzwarth, R.} \&
  \bibinfo{author}{Diddams, S.~A.}
\newblock \bibinfo{title}{Microresonator-based optical frequency combs}.
\newblock \emph{\bibinfo{journal}{Science}} \textbf{\bibinfo{volume}{332}},
  \bibinfo{pages}{555--559} (\bibinfo{year}{2011}).

\bibitem{iotti2001carrier}
\bibinfo{author}{Iotti, R.~C.} \& \bibinfo{author}{Rossi, F.}
\newblock \bibinfo{title}{Carrier thermalization versus phonon-assisted
  relaxation in quantum-cascade lasers: A monte carlo approach}.
\newblock \emph{\bibinfo{journal}{Applied Physics Letters}}
  \textbf{\bibinfo{volume}{78}}, \bibinfo{pages}{2902--2904}
  (\bibinfo{year}{2001}).

\bibitem{wittmann2008intersubband}
\bibinfo{author}{Wittmann, A.}, \bibinfo{author}{Bonetti, Y.},
  \bibinfo{author}{Faist, J.}, \bibinfo{author}{Gini, E.} \&
  \bibinfo{author}{Giovannini, M.}
\newblock \bibinfo{title}{Intersubband linewidths in quantum cascade laser
  designs}.
\newblock \emph{\bibinfo{journal}{Applied Physics Letters}}
  \textbf{\bibinfo{volume}{93}}, \bibinfo{pages}{141103}
  (\bibinfo{year}{2008}).

\bibitem{Henry1982Theory}
\bibinfo{author}{Henry, C.}
\newblock \bibinfo{title}{Theory of the linewidth of semiconductor lasers}.
\newblock \emph{\bibinfo{journal}{{IEEE} Journal of Quantum Electronics}}
  \textbf{\bibinfo{volume}{18}}, \bibinfo{pages}{259--264}
  (\bibinfo{year}{1982}).

\bibitem{green2008linewidth}
\bibinfo{author}{Green, R.~P.} \emph{et~al.}
\newblock \bibinfo{title}{Linewidth enhancement factor of terahertz quantum
  cascade lasers}.
\newblock \emph{\bibinfo{journal}{Applied Physics Letters}}
  \textbf{\bibinfo{volume}{92}}, \bibinfo{pages}{071106}
  (\bibinfo{year}{2008}).

\bibitem{jumpertz2016measurements}
\bibinfo{author}{Jumpertz, L.} \emph{et~al.}
\newblock \bibinfo{title}{Measurements of the linewidth enhancement factor of
  mid-infrared quantum cascade lasers by different optical feedback
  techniques}.
\newblock \emph{\bibinfo{journal}{{AIP} Advances}}
  \textbf{\bibinfo{volume}{6}}, \bibinfo{pages}{015212} (\bibinfo{year}{2016}).

\bibitem{piccardo2019frequency}
\bibinfo{author}{Piccardo, M.} \emph{et~al.}
\newblock \bibinfo{title}{Frequency-modulated combs obey a variational
  principle}.
\newblock \emph{\bibinfo{journal}{Physical Review Letters}}
  \textbf{\bibinfo{volume}{122}} (\bibinfo{year}{2019}).

\bibitem{Agrawal1988Population}
\bibinfo{author}{Agrawal, G.~P.}
\newblock \bibinfo{title}{Population pulsations and nondegenerate four-wave
  mixing in semiconductor lasers and amplifiers}.
\newblock \emph{\bibinfo{journal}{Journal of the Optical Society of America B}}
  \textbf{\bibinfo{volume}{5}}, \bibinfo{pages}{147} (\bibinfo{year}{1988}).

\bibitem{gaeta2019photonic}
\bibinfo{author}{Gaeta, A.~L.}, \bibinfo{author}{Lipson, M.} \&
  \bibinfo{author}{Kippenberg, T.~J.}
\newblock \bibinfo{title}{Photonic-chip-based frequency combs}.
\newblock \emph{\bibinfo{journal}{Nature Photonics}}
  \textbf{\bibinfo{volume}{13}}, \bibinfo{pages}{158--169}
  (\bibinfo{year}{2019}).

\bibitem{haensch2006nobel}
\bibinfo{author}{H\"{a}nsch, T.~W.}
\newblock \bibinfo{title}{Nobel lecture: Passion for precision}.
\newblock \emph{\bibinfo{journal}{Reviews of Modern Physics}}
  \textbf{\bibinfo{volume}{78}}, \bibinfo{pages}{1297--1309}
  (\bibinfo{year}{2006}).

\bibitem{hall2006nobel}
\bibinfo{author}{Hall, J.~L.}
\newblock \bibinfo{title}{Nobel lecture: Defining and measuring optical
  frequencies}.
\newblock \emph{\bibinfo{journal}{Reviews of Modern Physics}}
  \textbf{\bibinfo{volume}{78}}, \bibinfo{pages}{1279--1295}
  (\bibinfo{year}{2006}).

\bibitem{Keilmann2004Time}
\bibinfo{author}{Keilmann, F.}, \bibinfo{author}{Gohle, C.} \&
  \bibinfo{author}{Holzwarth, R.}
\newblock \bibinfo{title}{Time-domain mid-infrared frequency-comb
  spectrometer}.
\newblock \emph{\bibinfo{journal}{Optics Letters}}
  \textbf{\bibinfo{volume}{29}}, \bibinfo{pages}{1542} (\bibinfo{year}{2004}).

\bibitem{hugi2012mid}
\bibinfo{author}{Hugi, A.}, \bibinfo{author}{Villares, G.},
  \bibinfo{author}{Blaser, S.}, \bibinfo{author}{Liu, H.~C.} \&
  \bibinfo{author}{Faist, J.}
\newblock \bibinfo{title}{Mid-infrared frequency comb based on a quantum
  cascade laser}.
\newblock \emph{\bibinfo{journal}{Nature}} \textbf{\bibinfo{volume}{492}},
  \bibinfo{pages}{229--233} (\bibinfo{year}{2012}).

\bibitem{schwarz2019monolithic}
\bibinfo{author}{Schwarz, B.} \emph{et~al.}
\newblock \bibinfo{title}{Monolithic frequency comb platform based on interband
  cascade lasers and detectors}.
\newblock \emph{\bibinfo{journal}{Optica}} \textbf{\bibinfo{volume}{6}},
  \bibinfo{pages}{890} (\bibinfo{year}{2019}).

\bibitem{hillbrand2020inphase}
\bibinfo{author}{Hillbrand, J.} \emph{et~al.}
\newblock \bibinfo{title}{In-phase and anti-phase synchronization in a laser
  frequency comb}.
\newblock \emph{\bibinfo{journal}{Physical Review Letters}}
  \textbf{\bibinfo{volume}{124}} (\bibinfo{year}{2020}).

\bibitem{sterczewski2020frequency}
\bibinfo{author}{Sterczewski, L.~A.}, \bibinfo{author}{Frez, C.},
  \bibinfo{author}{Forouhar, S.}, \bibinfo{author}{Burghoff, D.} \&
  \bibinfo{author}{Bagheri, M.}
\newblock \bibinfo{title}{Frequency-modulated diode laser frequency combs at 2
  $\upmu$m wavelength}.
\newblock \emph{\bibinfo{journal}{{APL} Photonics}}
  \textbf{\bibinfo{volume}{5}}, \bibinfo{pages}{076111} (\bibinfo{year}{2020}).

\bibitem{singleton2018evidence}
\bibinfo{author}{Singleton, M.}, \bibinfo{author}{Jouy, P.},
  \bibinfo{author}{Beck, M.} \& \bibinfo{author}{Faist, J.}
\newblock \bibinfo{title}{Evidence of linear chirp in mid-infrared quantum
  cascade lasers}.
\newblock \emph{\bibinfo{journal}{Optica}} \textbf{\bibinfo{volume}{5}},
  \bibinfo{pages}{948} (\bibinfo{year}{2018}).

\bibitem{hillbrand2018tunable}
\bibinfo{author}{Hillbrand, J.}, \bibinfo{author}{Jouy, P.},
  \bibinfo{author}{Beck, M.} \& \bibinfo{author}{Faist, J.}
\newblock \bibinfo{title}{Tunable dispersion compensation of quantum cascade
  laser frequency combs}.
\newblock \emph{\bibinfo{journal}{Optics Letters}}
  \textbf{\bibinfo{volume}{43}}, \bibinfo{pages}{1746} (\bibinfo{year}{2018}).

\bibitem{herr2013temporal}
\bibinfo{author}{Herr, T.} \emph{et~al.}
\newblock \bibinfo{title}{Temporal solitons in optical microresonators}.
\newblock \emph{\bibinfo{journal}{Nature Photonics}}
  \textbf{\bibinfo{volume}{8}}, \bibinfo{pages}{145--152}
  (\bibinfo{year}{2013}).

\bibitem{aranson2002ginzburg}
\bibinfo{author}{Aranson, I.~S.} \& \bibinfo{author}{Kramer, L.}
\newblock \bibinfo{title}{{The world of the complex Ginzburg-Landau equation}}.
\newblock \emph{\bibinfo{journal}{Reviews of Modern Physics}}
  \textbf{\bibinfo{volume}{74}}, \bibinfo{pages}{99--143}
  (\bibinfo{year}{2002}).

\bibitem{columbo2020unifying}
\bibinfo{author}{Columbo, L.} \emph{et~al.}
\newblock \bibinfo{title}{Unifying frequency combs in active and passive
  cavities: Temporal solitons in externally-driven ring lasers}.
\newblock \emph{\bibinfo{journal}{arXiv:2007.07533v2}}  (\bibinfo{year}{2020}).

\bibitem{bastard1988wave}
\bibinfo{author}{Bastard, G.}
\newblock \emph{\bibinfo{title}{Wave mechanics applied to semiconductor
  heterostructures}} (\bibinfo{publisher}{Les Editions de Physique},
  \bibinfo{year}{1988}).

\bibitem{terazzi2008sequential}
\bibinfo{author}{Terazzi, R.}, \bibinfo{author}{Gresch, T.},
  \bibinfo{author}{Wittmann, A.} \& \bibinfo{author}{Faist, J.}
\newblock \bibinfo{title}{Sequential resonant tunneling in quantum cascade
  lasers}.
\newblock \emph{\bibinfo{journal}{Physical Review B}}
  \textbf{\bibinfo{volume}{78}} (\bibinfo{year}{2008}).

\bibitem{terazzi2010density}
\bibinfo{author}{Terazzi, R.} \& \bibinfo{author}{Faist, J.}
\newblock \bibinfo{title}{A density matrix model of transport and radiation in
  quantum cascade lasers}.
\newblock \emph{\bibinfo{journal}{New Journal of Physics}}
  \textbf{\bibinfo{volume}{12}}, \bibinfo{pages}{033045}
  (\bibinfo{year}{2010}).

\bibitem{jirauschek2014modeling}
\bibinfo{author}{Jirauschek, C.} \& \bibinfo{author}{Kubis, T.}
\newblock \bibinfo{title}{Modeling techniques for quantum cascade lasers}.
\newblock \emph{\bibinfo{journal}{Applied Physics Reviews}}
  \textbf{\bibinfo{volume}{1}}, \bibinfo{pages}{011307} (\bibinfo{year}{2014}).

\bibitem{harrison2003quantum}
\bibinfo{author}{Harrison, P.} \& \bibinfo{author}{Valavanis, A.}
\newblock \emph{\bibinfo{title}{Quantum Wells, Wires and Dots: Theoretical and
  Computational Physics of Semiconductor Nanostructures, Fourth Edition}}
  (\bibinfo{publisher}{Wiley}, \bibinfo{year}{2016}).

\bibitem{cuda}
\bibinfo{author}{NVIDIA}, \bibinfo{author}{Vingelmann, P.} \&
  \bibinfo{author}{Fitzek, F.~H.}
\newblock \bibinfo{title}{Cuda, release: 10.2.89} (\bibinfo{year}{2020}).
\newblock \urlprefix\url{https://developer.nvidia.com/cuda-toolkit}.

\end{thebibliography}

%************** Acknowledgement **************
%\vspace{-0.2cm}
\section*{Acknowledgements}
\footnotesize
\setstretch{1.}
This project has received funding from the European Research Council (ERC) under the European Union’s Horizon 2020 research and innovation programme (Grant agreement No. 853014).
We acknowledge discussions with A.~Belyanin and Y.~Wang on the origin of the linewidth enhancement factor and discussions with G.~Bastard on the higher-order scattering mechanisms.

\section*{Author contributions}
\footnotesize
\setstretch{1.}
N.O. derived the theoretical framework. N.O. and B.S. implemented the numerical framework and performed the simulations. S.D.C and J.H. analysed the results in the experimental context. B.S. supervised the work. N.O. wrote the manuscript with editorial input from S.D.C., J.H. and B.S. All authors contributed to the analysis and discussion of the paper.
\end{document}